# SERS Mechanism on Graphene

**V. P. Chelibanov[a], S. A. Ktitorov[b], A. M. Polubotko[b*], Yu. A. Firsov[b]**

[a] State University of Information Technologies, Mechanics and Optics, 197101 Saint Petersburg, RUSSIA
[b] A.F. Ioffe Physico-Technical Institute, The Department of Dielectrics and Semiconductors, The Sector of the Theory of Semiconductors and Dielectrics,194021 Saint Petersburg, RUSSIA
E-mail: alex.marina@mail.ioffe.ru

**Abstract**

The paper presents briefly some general points of the theory of Surface Enhanced Raman Scattering on metals, semiconductors, dielectrics and graphene. It is pointed out that the reason of SERS on graphene and some other 2D materials is not the "chemical enhancement mechanism" as it is widely accepted in literature, but ripples, which are necessarily present on a graphene surface and are analogues of the surface roughness, which is a reason of SERS. In addition, it is indicated that the quadrupole interaction, which arises in strong SERS is practically absent in the system with graphene. The results are confirmed by the analysis of the SERS, usual Raman and IR spectra of phthalocyanine molecule and by some other experimental data.

## 1. Introduction

Surface Enhanced Raman Scattering (SERS) is a well known effect, which is well investigated experimentally. Usually researchers consider that the reason of SERS is surface plasmons and a "chemical enhancement". This opinion is widespread in literature and is given even in monographs [1, 2] for example. One considers that the surface plasmons result in the enhancement factor $G_{electr.} \sim 10^4$ due to enhancement of electromagnetic field, while the second, "chemical mechanism", associated with the direct interaction of the molecules of the first adsorbed layer with the substrate, gives the enhancement factor $G_{chem} \sim 10^2$. At present SERS is observed on graphene and is named as Graphene Enhanced Raman Scattering (GERS) [3]. The enhancement factor for GERS is $G_{GERS} \sim 10^2$ that results in opinion that its reason is the "chemical enhancement mechanism". As it is well known, the reason of SERS is a surface roughness. Therefore, in literature and in our works there is another opinion that the reason of SERS is so-called rod effect, which is associated with the enhancement of the electromagnetic field near the places of the surface with a very large positive curvature. In addition, we consider that it is necessary to take into account not only the dipole, but the quadrupole light-molecule interaction, which can be very significant in SERS.

In [4-6] we demonstrated that using the concept of the rod effect, the additional enhancement in the first layer can be explained by a pure electrodynamical mechanism and its appearance is associated with a strong change of the electric field in space that results in a strong difference of the enhancement in the first and in the second layers of adsorbed molecules. Therefore, the value $"G"_{chem} \sim 10^2$ can be explained completely using this idea. Here in the paper we would like to present our main ideas about the enhancement mechanism in SERS and to point out a main possible reason of GERS.

## 2. Main properties of an electromagnetic field near a rough metal surface.

The main property of the electromagnetic field near a rough surface is its strong spatial inhomogeneity. As an example we can consider its properties near a model of the rough surface - a strongly jagged lattice with a regular triangular profile of the eshelett type (Fig. 1). The H-polarized electromagnetic field above this lattice can be represented in the form

$$\mathbf{E} = \mathbf{E}_{inc} + \mathbf{E}_{surf..sc} \qquad (1)$$

where

$$\mathbf{E}_{inc} = \mathbf{E}_{0,inc} e^{-ik_0 \cos\theta_0 z + ik_0 \sin\theta_0 y}$$
$$|\mathbf{E}_{0,inc}| = 1 \qquad (2)$$

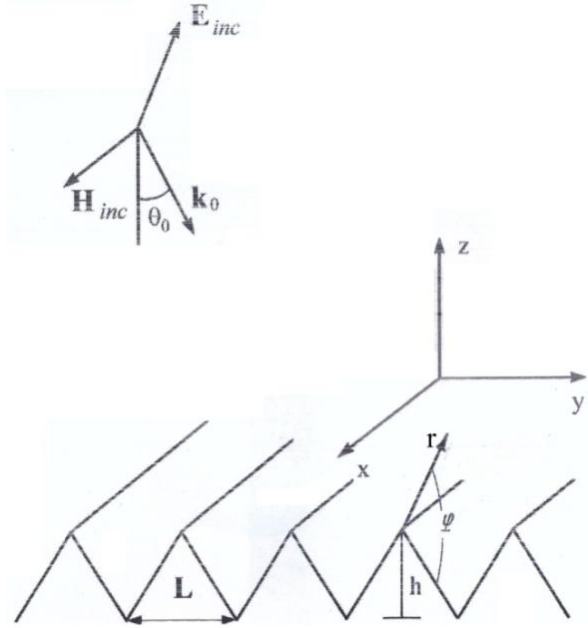

Fig. 1. Diffraction of a plane H-polarized electromagnetic wave on the lattice of the eshelett type. $r$ and $\varphi$ are the radial and the angle coordinates of the cylindrical coordinate system near a wedge.

$\theta_0$ - is the angle of incidence, $k_0 = |\mathbf{k_0}|$ is the module of the wave vector of the incident field in a free space,

$$\mathbf{E}_{surf.sc.} = \sum_{n=-\infty}^{+\infty} \mathbf{g}_n e^{i\alpha_n y + i\gamma_n z}, \qquad (3)$$

$$\alpha_n = \frac{2\pi n + k_0 \sin\theta_0 L}{L}, \qquad (4)$$

$$\gamma_n = \sqrt{k_0^2 - \alpha_n^2} \quad . \tag{5}$$

Here $\mathbf{g}_n$ are the amplitudes, $n$ is the number of a spatial harmonic. For the period of the lattice $L \ll \lambda$ the spatial harmonic with $n=0$ is a direct reflected wave, while all others are inhomogeneous plane waves strongly localized near the surface. The maximum localization size has the harmonic with $n=1$. All others are localized considerably stronger. The exact solution of the diffraction problem on the lattice reduces to determination of the coefficients $\mathbf{g}_n$. The main specific feature of the surface field is a steep or singular increase of the electric field near the wedges of the lattice or so-called rod effect. This type of behavior is independent on a particular surface profile. It is determined only by existence of sharp wedges. Besides it is independent on the dielectric properties of the lattice and exists in lattices with any dielectric constants that differ from the dielectric constant of vacuum. In addition such behavior must presence on a graphene lattice of the above described profile, when this lattice is only of one or a few layers of carbon.

For simplicity let us consider the behavior of the H-polarized electromagnetic field with a normal incidence. Here we shall consider the electric field near the wedge of the lattice (Figure 2). It can be a metallic, semiconductor, dielectric, or "graphene" wedge. The main property of the behavior of the electric field near the wedge is the necessity to tend to infinity, when one moves to its top. This property follows from the fact that the direction of the electric field on the top is not defined and is arbitrary. Such situation can be realized only in two cases, when the electric field is equal to zero and when it is infinity. It is obvious that the first case can be realized only for some special conditions, while usually we have the second case. This means that for any lattice: metal, semiconductor, dielectric or graphene there must be a huge enhancement of the electric field on the top of the wedge. Considering the normal and tangential components of the electric field near the top, we can obtain for the derivatives of these components along the $y$ axis

$$\frac{\partial \mathbf{E}_n}{\partial y} = \frac{\mathbf{E}_n(a) - \mathbf{E}_n(b)}{\Delta y} \to \infty , \tag{6}$$
$$\Delta y \to 0$$

$$\frac{\partial \mathbf{E}_\tau}{\partial y} = \frac{\mathbf{E}_\tau(a) - \mathbf{E}_\tau(b)}{\Delta y} \to \infty , \tag{7}$$
$$\Delta y \to 0$$

when one moves along the sides of the wedge to the top. This means that both the normal and the tangential components of the electric field must be enhanced near this point. The only exclusion arises when one considers an ideally conductive metal wedge, when the tangential component $\mathbf{E}_\tau \equiv 0$ on the surface. In the cylindrical coordinate system for this case [2], the behavior of the $E_r = \mathbf{E}_\tau$ and $E_\varphi$ components near the top of the wedge can be expressed as

$$E_r = -|E_{0,inc}|C_0 \left(\frac{l_1}{r}\right)^\beta \sin(\lambda_1 \varphi)$$

$$E_\varphi = -|E_{0,inc}|C_0 \left(\frac{l_1}{r}\right)^\beta \cos(\lambda_1 \varphi)$$

(8)

where $C_0$ is some numerical coefficient, $(l_1 = L \text{ or } h)$ is a characteristic size of the lattice

$$\lambda_1 = \pi/(2\pi - \alpha) \tag{9}$$

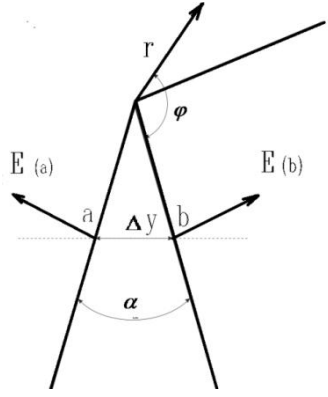

Fig. 2. The behavior of the electric field for normal incidence of the H-polarized incident field near the wedge of the lattice. For simplicity the figure corresponds to an ideally conductive wedge. For the semiconductor, dielectric and graphene wedge the electric field on the boundaries can have not a normal direction.

$\alpha$ is the wedge angle for an ideally conductive wedge,

$$\beta = 1 - \lambda_1 = \frac{\pi - \alpha}{2\pi - \alpha} \tag{10}$$

The specific feature of the field behavior (8) is appearance of the singularity $(l_1/r)^\beta$, which describes geometrical nature of the field enhancement. It determines the following behavior of the coefficients $\mathbf{g}_n$ in expression (3):

$$|\mathbf{g_n}| \sim |n|^{\beta-1}. \tag{11}$$

Indeed, substitution of (11) into (3) gives

$$\sum_{\substack{n=-\infty \\ n \neq 0}}^{+\infty} |n|^{\beta-1} e^{2\pi|n|z/L} \sim 2\int_0^\infty t^{\beta-1} e^{-2\pi t/L} dt \sim 2\left(\frac{L}{2\pi z}\right)^\beta \tag{12}$$

For the wedge angles changing in the interval $0 < \alpha < \pi$ the parameter $\beta$ varies within the interval $0 < \beta < 1/2$ and the coefficients $\mathbf{g_n}$ slowly decrease as $n$ increases. Thus, the singular behavior of the field arises because of specific summation of the surface waves at the top of the wedge.

From (8) one can see that the zero equality for the case of an ideally conductive wedge is provided by the factor $\sin \lambda_1 \varphi$, which is equal to zero on the boundaries of the wedge. However, for the dielectric wedge the boundary conditions are another (they are the equality of tangential components of the electric and magnetic field at the boundaries) and therefore $E_r = \mathbf{E}_\tau \neq 0$. This component will be singular and tends to infinity also, when one moves to the top of the wedge

$$E_r, E_\varphi \sim \left(\frac{l_1}{r}\right)^\beta \tag{13}$$

here,

$$\beta = 1 - \tau \tag{14}$$

and differs from the case of ideally conductive wedge. The $\tau$ value is defined from the solution of the following transcendent equation [7-8]

$$F_1(\tau) + \frac{F_2(\tau)}{\varepsilon} + \frac{F_1(\tau)}{\varepsilon^2} = 0 \tag{15}$$

where $\varepsilon$ is a dielectric constant of the wedge

$$F_1(\tau) = \sin(2\pi - \alpha)\tau \times \sin \alpha \tau \tag{16}$$

$$F_2(\tau) = 2[1 - \cos(2\pi - \alpha)\tau \times \cos \alpha \tau] \tag{17}$$

Thus, there is an enhancement both of the $\mathbf{E}_n$ and $\mathbf{E}_\tau$ components for the dielectric wedge. The same situation must be for the graphene wedge since the condition $\mathbf{E}_\tau = 0$ is not valid for this case also because it is only of one atomic layer width and the field penetrates through graphene.

In the region of a three-dimensional roughness of the cone or tip type the formula for estimation of the field has an approximate form for all types of materials

$$E_r \sim |\mathbf{E}_{0,inc}| C_0 \left(\frac{l_1}{r}\right)^\beta, \tag{18}$$

where $\beta$ depends on the cone angle and the dielectric properties of the material and varies within the interval $0 < \beta < 1$. Using formulae (8) and (18) one can note a very important property: a strong spatial variation of the field. For example

$$\frac{1}{E_r}\frac{\partial E_r}{\partial r} \sim \left(\frac{\beta}{r}\right) \tag{19}$$

can be significantly larger than the value $2\pi/\lambda$, which characterizes variation of the electric field in a free space. If one considers more realistic models of the rough surface, when the wedges have a finite curvature at the top, than there is a strong enhancement of the $\mathbf{E}_n$ component of the electric field at places of the surface with a large curvature for the metal, semiconductor, dielectric or graphene lattices, while the tangential component $\mathbf{E}_\tau$ is comparable with the amplitude of the incident field for an ideally conductive case and must be enhanced for the case of the semiconductor, dielectric or graphene. Thus the peculiarity, indicated above is principal and there is a great difference between the cases of a metal and of a semiconductor, dielectric and graphene.

1. Some notions about the "chemical enhancement mechanism"

As it has been indicated above at present, it is considered in literature that the reason of GERS is so-called "chemical enhancement", or the one, associated with the change of the electron structure of molecules due to their direct interaction with the surface in the first adsorbed layer [1]. However, as it was demonstrated in [4-6], there is no the "chemical enhancement" and the one in SERS is associated with very strong change of the electric field, when one moves away from the surface. Therefore, the electric field and its derivatives in the first layer of adsorbed molecules near the places with a very large positive curvature differ strongly from the ones in the second and upper layers. This result follows from consideration of the electromagnetic field (8,18) near the top of the roughness of the wedge, or a tip form (Fig. 3). Here we present the result for the metal wedge. In this case there arises strong dipole and quadrupole interaction, which are associated with the enhancement of the electric field and its derivatives with equal indices. The corresponding relations for the first and the second layers of adsorbed molecules follow from the Figure 3 and are as follows

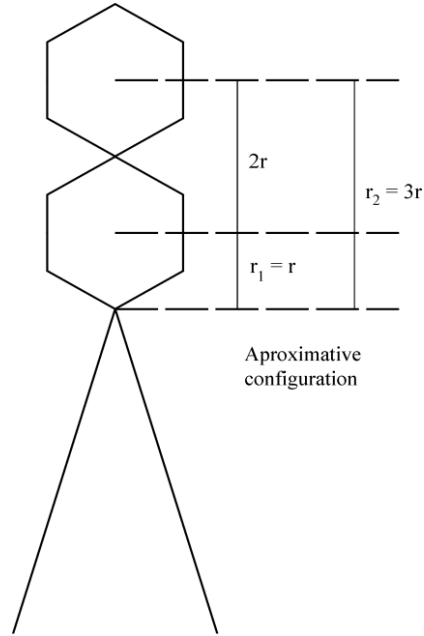

Fig. 3. The relation of the electric field, or its derivatives in the first and the second layer explain the "chemical enhancement" completely.

$$\frac{E_r^4(r_1)}{E_r^4(r_2)} \cong \left(\frac{r_2}{r_1}\right)^{4\beta} \sim (3)^{4\beta} \sim 100 \qquad (20)$$

$$\left(\frac{\partial E_r}{\partial r}\right)^4_{r=r_1} \bigg/ \left(\frac{\partial E_r}{\partial r}\right)^4_{r=r_2} \cong \left(\frac{r_2}{r_1}\right)^{4+4\beta} = (3)^{4\beta+4} \sim 100-1000 \tag{21}$$

**O**ne can see that the values obtained are able to explain the "chemical enhancement", or the first layer effect [1] completely. This result in principle was confirmed experimentally for GERS in [3], where the authors demonstrated, that the larger the distance of the molecules from the graphene surface, the lower the enhancement. The most enhancement arises for the molecules, which are in the first layer of adsorbed molecules. In addition it is necessary to note that in case the "chemical enhancement" would be associated with the direct interaction of molecules with a substrate, it should be observed on a perfectly flat, or in case of a single surface also. However, as it was demonstrated by A. Campion, [9] for example, there is no any enhancement on such surfaces. Therefore, we consider that the idea of the "chemical enhancement" is completely incorrect and an explanation of GERS within this concept is not correct also.

## 2. About the GERS mechanism

Graphene is not ideally flat because of so-called ripples (Fig. 4). Therefore, there must be an electromagnetic field, which is localized near the graphene surface such as on the above considered lattices. Since graphene has a very large conductivity and electron mobility (the latter is significantly larger than in silver, for example), graphene is a sufficiently strong perturbation for electromagnetic waves. Therefore, both the normal and the tangential components of the electric field must be enhanced near the tops of the ripples (Fig. 5). The fact that the ripples are rather smooth and graphene is only a single-atomic layer and not a massive metal, results in a weak enhancement of the electric field and of GERS with the enhancement factor $G_{GERS} \sim 10^2$. One should recall that the enhancement coefficient in usual SERS is $G_{SERS} \sim 10^6$. The fact that the ripples are the reason of GERS can be confirmed by the following experimental facts. In accordance with the results of [10], the enhancement in GERS becomes lower, when we increase the number of layers of graphene. We consider that this result is associated with the decrease of the height of the ripples for multilayer graphene.

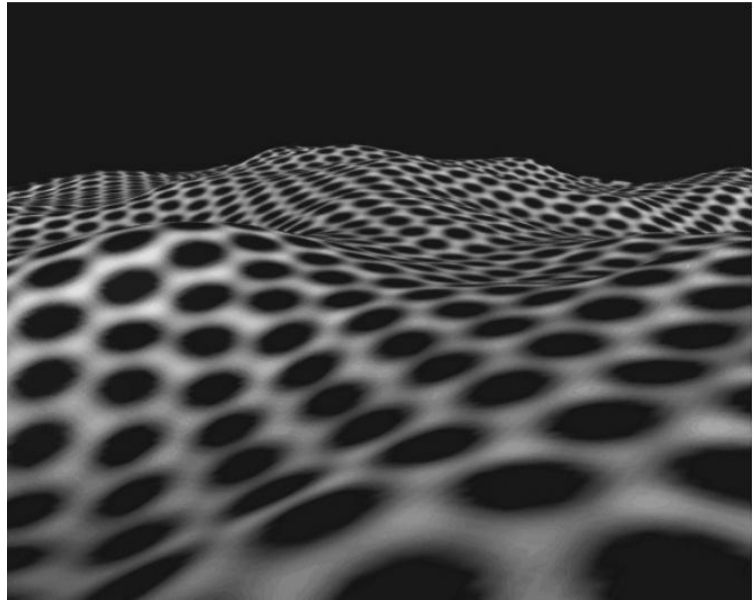

Fig. 4. Real graphene is not an ideal plane, but there are so-called ripples.

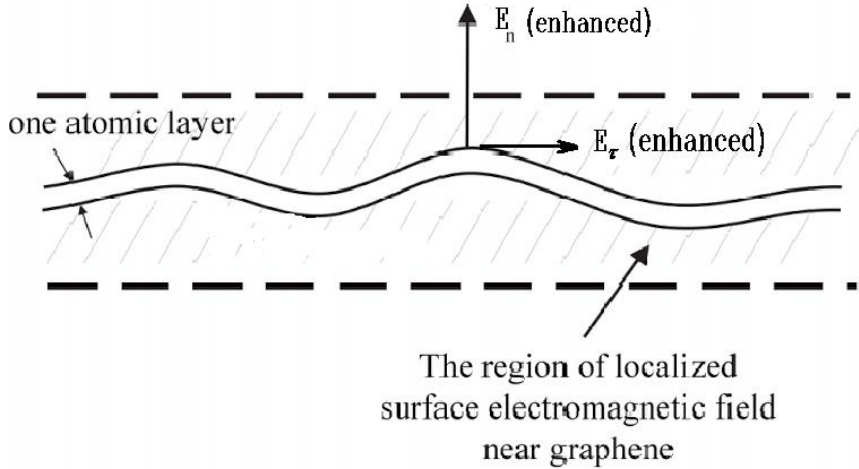

Fig. 5. Enhancement of the electric field near the tops of the ripples.

The interaction between the graphene layers increases rigidity in the direction, which is perpendicular to its plane that results in the decrease of the height of the ripples. They become smoother and, therefore, the enhancement decreases.

As it is well known from the dipole-quadrupole SERS theory [2], the quadrupole interaction is effective in the usual strong SERS and causes appearance of forbidden lines in symmetrical molecules. However, because of a weak enhancement of the electric field and its derivatives on grapheme, the quadrupole interaction must be weak or absent at all. This result can be well confirmed by the analysis of the GERS spectra of symmetrical molecules. The GERS spectra of phthalocianine molecule ($H_2Pc$) were observed and investigated in [10]. This molecule is very large and consists of 58 atoms. Therefore, the mean interval between neighbor vibrational lines is something about 15 $cm^{-1}$ that is comparable or less of the calculation errors of wavenumbers. Therefore, we believe that the direct assignment of the experimental GERS spectra on the base of calculations made in [11] is impossible. However, in [11] the authors experimentally measured the position of the IR and usual Raman lines of $H_2Pc$. Taking into account that there is a rule of mutual exclusion in the IR and Raman spectra one can convince that practically all GERS lines strongly correspond to their usual Raman lines, while almost all IR lines do not correspond to the GERS lines except the ones at 718, 1008,1185, 142,1538 and 1614 $cm^{-1}$ (Table 1). This result however is associated with the fact that the wavenumbers of these IR lines practically coincide with usual Raman lines that indicates that these pairs are nearly degenerated. Therefore, we cannot conclude practically that there are the lines, which are forbidden in the GERS spectra. From very approximate results obtained in [11] one can conclude that all the GERS lines of phthalocyanine can be assigned to $A_g, B_{1g}, B_{2g}$ and $B_{3g}$ irreducible representations of the $D_{2h}$ symmetry group, which describes the symmetry properties of the molecule. This result indicates in principle on the enhancement of both the normal and tangential components of the electric field on the graphene surface since this result correspond to the dipole-dipole type of the scattering only.

Table 1. Experimental wavenumbers for the SERS, usual Raman and IR spectra of $H_2Pc$

| Experimental wavenumbers and qualitative relative intensity for SERS of $H_2Pc$ adsorbed on graphene in $cm^{-1}$ | Experimental wavenumbers for Raman scattering on $H_2Pc$ in $cm^{-1}$ | Experimental wavenumbers for IR absorption in $H_2Pc$ in $cm^{-1}$ |
|---|---|---|
| 681 vs | 680 | |
| 723 vs | 721 | 718 double s |
| 766 vw | | |
| 796 m | 795 | |
| 890 w | | |
| 1007 vw | 1009 | 1008 |
| 1024 m | 1024 | |
| 1082 m | 1082 | |
| 1107 s | 1107 | |
| 1141 s | 1138 | |
| 1182 s | 1179 | 1185 |
| 1228 m | 1227 | |
| 1312 s | 1310 | |
| 1340 s | 1340 | |
| 1405 vw | 1398 | 1402 |
| 1428 m | 1425 | |
| 1450 m | 1448 | |
| 1479 w | | |
| 1512 m | 1513 | |
| 1543 vs | 1541 | 1538 |
| 1617 vw | 1618 | 1614 |

The actual absence of the forbidden lines with $B_{1u}, B_{2u}$ and $B_{3u}$ symmetry, which might arise due to the dipole-quadrupole scattering type [2] also indicates that the spectrum is determined practically only by the enhancement of the electric field and by the dipole interaction, while the quadrupole interaction in this system is very weak and practically does not manifest in the GERS spectra. One should note that there is SERS on some other 2D materials: $MoS_2$ and $h$-$BN$ [12]. In addition, it was pointed out in [13] that there are ripples on $MoS_2$. The fact that these materials are semiconductors and dielectrics, but not a zero-gap semiconductor as graphene can explain the fact that SERS on these materials is weaker. Since graphene has a very high conductivity and mobility, it provides stronger inhomogeneity in space and, therefore, stronger SERS. SERS on $MoS_2$ and $h$-$BN$ requires more careful investigation with their specific properties taking into account. Therefore, here we do not consider this topic in detail. However, the fact that there are experimental data indicating existence of ripples on $MoS_2$ strongly confirms our point of view.

## 3. Conclusion

Thus, from our opinion the reason of GERS is not the "chemical mechanism" but the ripples, which are necessarily present on the graphene surface. In spite of a very large conductivity and mobility of electrons in graphene, the fact that the pure graphene is only a single-atomic layer indicates that the inhomogeneity of this system is not so strong as for a massive metal like silver. In addition, the ripples are sufficiently smooth that results in a sufficiently weak enhancement of the electric field. In principle, both the normal and the tangential components of the electric field are enhanced near the graphene surface. The fact that the enhancement of the electric field is weak is confirmed by the low value of the GERS enhancement coefficient and from the analysis of the GERS spectrum of phthalocyanine. As it follows from these experimental data, the quadrupole interaction, which always presents in strong SERS is very weak and is practically absent. It should be noted that our point of view is confirmed by the fact that SERS exists on other 2D materials - $MoS_2$ and $h\text{-}BN$ and by the fact that there are ripples on $MoS_2$.